\documentclass[a4paper,11pt]{article}

\usepackage{a4wide}
\usepackage[english]{babel}
\usepackage{amsmath,amssymb,amsfonts,amsthm}
\usepackage{color}
\usepackage{graphicx}
\usepackage{calc}
\usepackage{subfigure}

\newcommand{\starcom}[2]{[#1\stackrel{\star}{,}#2]}

\title{Spacetime~Noncommutativity in Models~with~Warped~Extradimensions}
\author{
  Thorsten Ohl\thanks{e-mail: \texttt{ohl@physik.uni-wuerzburg.de}}\qquad
  Alexander Schenkel\thanks{e-mail: \texttt{aschenkel@physik.uni-wuerzburg.de}}\qquad
  Christoph F.~Uhlemann\thanks{e-mail: \texttt{uhlemann@physik.uni-wuerzburg.de}} \\
  \hfill\\
  Institut f\"ur Theoretische Physik und Astrophysik\\
    Universit\"at W\"urzburg\\
  Am Hubland, 97074 W\"urzburg, Germany}

\begin{document}
\maketitle
\begin{abstract}
We construct consistent noncommutative (NC) deformations of the
Randall-Sundrum spacetime that solve the NC Einstein equations with a
non-trivial Poisson tensor depending on the fifth coordinate.
In a class of these deformations where the Poisson tensor is
exponentially localized on one of the branes (the NC-brane), we study
the effects on bulk particles in terms of Lorentz-violating operators
induced by NC-brane interactions.  We sketch two models in which
massive bulk particles mediate NC effects to an almost-commutative
SM-brane, such that observables at high energy colliders are
enhanced with respect to low energy and astrophysical observables.
\end{abstract}

\section{Introduction}
The noncommutativity~(NC) of spacetime coordinates is a generic
prediction of theories of quantum gravity, including string theory.
If the corresponding energy scale is low enough, effects of such a
NC could be observed in collider
experiments~\cite{Hewett:2000zp,Hinchliffe:2002km}.  Indeed, a
consistent effective theory of a NC extension of the
standard model can be
constructed~\cite{Jurco:2001rq,Calmet:2001na,Melic:2005fm,Melic:2005am}
and detailed calculations of collider effects have been
performed~(see, e.\,g., \cite{Ohl:2004tn,Alboteanu:2006hh,Alboteanu:2007bp}).
However, for this to be experimentally feasible, some rather stringent
constraints from low energy precision experiments and astrophysical
observations~(see~\cite{Hinchliffe:2002km} and references therein)
have to be avoided.

Theories with extra spacetime dimensions provide an attractive
scenario for reducing the scale of quantum gravity from the
four-dimensional Planck mass and are motivated by string theory.
Unfortunately, if naive dimensional analysis can be trusted, low energy
and astrophysical bounds push the NC scale far above the range accessible
to terrestrial colliders~\cite{Hinchliffe:2002km}.
Therefore we need a mechanism for enhancing
the collider scale NC effects relative to the low energy NC effects.

The aim of this paper is twofold: we develop a consistent description
of NC in models with extra spacetime dimensions and we
show how a combination of brane-localized NC, loop factors, moderate tuning
of parameters and resonance enhancement can lead to amplified NC
effects at the TeV-scale.

In section~\ref{sec:NCRS} we describe consistent deformations of the
Randall-Sundrum background and in section~\ref{sec:bulk} we study the
effects of these deformations on particles propagating in the bulk.
We introduce two exemplary models in section~\ref{sec:models}. While these
models are not intended as complete models for physics beyond the
standard model, we can use them to sketch some
phenomenological options in section~\ref{sec:pheno}, before concluding
in section~\ref{sec:concl}.

\section{Noncommutative Randall-Sundrum background}
\label{sec:NCRS}
\subsection{Kinematical setup}
In this paper we consider NC deformations of Randall-Sundrum
scenarios, which effectively reduce to a Moyal-Weyl deformation from
the four-dimensional perspective.  
To fix our notation, we briefly introduce the classical setup known as
RS1~\cite{Randall:1999ee}. 
Starting with the orbifold $\mathbb R^{1,3}\times S^1/\mathbb Z_2$
with two branes placed at the orbifold fixed-points, one obtains as a
solution of the classical 5D Einstein equations a slice of $\text{AdS}_5$
with fixed relations among the brane and bulk cosmological constants. 
We use the coordinates~$x^M=(x^\mu,y)$ in which the RS metric takes the form
\begin{flalign}
\label{eqn:metric}
 ds^2 = e^{-2kR|y|} \eta_{\mu\nu} dx^\mu \otimes dx^\nu - R^2 dy\otimes dy~.
\end{flalign}
The two branes of opposite tension are located at~$y=0$ and~$y=\pi$
and we refer to them as UV and IR brane, respectively.  

We now construct a NC RS spacetime.
Consider the following Poisson tensor on the RS background
\begin{flalign}
\label{eqn:poisson}
 \Theta^{\mu\nu}(y) = \vartheta(|y|) \omega^{\mu\nu} ~,\quad \Theta^{5\mu} = 0~,
\end{flalign}
where~$\omega^{\mu\nu}$ is a general constant antisymmetric matrix and for consistency with the orbifold symmetry $\vartheta$ depends on the modulus $|y|$ only.
We briefly show how to construct a $\star$-product realizing the
associated $\star$-commutation relations
\begin{flalign}
\label{eqn:comrelations}
 \starcom{x^\mu}{x^\nu} := x^\mu\star x^\nu - x^\nu\star x^\mu
     = i \lambda \Theta^{\mu\nu}(y)~,\quad \starcom{x^\mu}{y}=0~.
\end{flalign}
For this, we take a (possibly degenerate) constant
matrix~$T_a^{~\mu}$ and the canonical Poisson
tensor~$\theta_\mathrm{can}^{ab}$ of rank 4 in Darboux form, i.e.~
\begin{flalign}
\label{eqn:canonical}
 \theta_\mathrm{can}^{ab} = \begin{pmatrix}
         0 & 1 & 0 & 0 \\
	-1 & 0 & 0 & 0 \\
	 0 & 0 & 0 & 1 \\
	 0 & 0 & -1 & 0
        \end{pmatrix}~.
\end{flalign}
We can always find a~$T_a^{~\mu}$, such that $ \omega^{\mu\nu} =
T_a^{~\mu}T_b^{~\nu}\theta_\mathrm{can}^{ab}$ holds true.
We take the set of commuting vector fields~$\lbrace X_a\rbrace$ defined by 
\begin{flalign}
\label{eqn:vectorfields}
  X_1=T_1^{~\mu}\partial_\mu~,\quad
  X_2=\vartheta(|y|)T_2^{~\mu}\partial_\mu~,\quad
  X_3=T_3^{~\mu}\partial_\mu~,\quad
  X_4=\vartheta(|y|)~T_4^{~\mu}\partial_\mu~,
\end{flalign}
and construct a $\star$-product of RJS-type~\cite{Reshetikhin:1990ep,Jambor:2004kc}
\begin{flalign}
\label{eqn:RJSproduct}
 (f \star g)(x^M) := f(x^M)~\exp\Bigl(\frac{i\lambda}{2}
     \overleftarrow{X_a}\theta_\mathrm{can}^{ab}\overrightarrow{X_b}\Bigr)~g(x^M)~,
\end{flalign}
leading to the desired commutation relations~(\ref{eqn:comrelations}).
Note that, despite the appearance of non-differentiable functions depending on $|y|$, 
(\ref{eqn:RJSproduct}) is a well-defined $\star$-product since the derivatives 
act on the coordinates $x^\mu$ alone.

The $\star$-product (\ref{eqn:RJSproduct}) is constructed in such a way that~$X_1$ and~$X_3$
are Killing vector fields.  As shown
in~\cite{Schupp:2009pt,Ohl:2009pv,Aschieri:2009qh}, these so-called
semi-Killing $\star$-products (more precisely, the relevant structure is
the Drinfel'd twist) have the remarkable property that the
NC Einstein tensor reduces to the undeformed one.  
Furthermore, the classical stress-energy tensor of the RS model is a natural
choice in the NC setup, since potential
contributions from the 
$\star$-products drop out due to 4D translation 
symmetry and the semi-Killing property of the $\star$-product.
Note that the brane-localized terms cause no obstructions,
since the $\star$-products act on $x^\mu$ alone.
Therefore, our NC RS background is a solution of the NC Einstein equations proposed
in~\cite{Aschieri:2005yw,Aschieri:2005zs} with the classical stress-energy tensor as a source. 
We refer to~\cite{Schupp:2009pt,Ohl:2009pv,Aschieri:2009qh,Asakawa:2009yb,Stern:2009id}
for more on NC gravity solutions.

This completes the description of our kinematical setup.  Next, we
want to face the question of how to choose the up to now arbitrary
function~$\vartheta(|y|)$.  From the NC gravity perspective we have
already chosen a preferred $\star$-product of semi-Killing type, such
that the classical background solves the NC Einstein equations.  In
this context we could further specialize to a constant~$\vartheta(|y|)$
to obtain a $\star$-product constructed entirely from Killing vector
fields.  However, to determine~$\vartheta(|y|)$ we will use a different,
string theory inspired perspective, providing additional prospects for
model-building.

\subsection{String theory inspired noncommutativity\label{sec:stringNC}}
String theory with a non-vanishing background for the $B$-field
arising in the massless closed-string spectrum is known to lead to
NC effects in string scattering amplitudes (see
e.g.~\cite{Douglas:1997fm,Schomerus:1999ug,Seiberg:1999vs}). Additionally,
it was shown by Seiberg and Witten~\cite{Seiberg:1999vs} that a
NC (super) Yang-Mills theory can be obtained in the low
energy limit.  The $B$-field background defines the Poisson tensor in
the $\star$-product via
\begin{flalign}\label{eqn:PoissonBfield}
\lambda \Theta^{\mu\nu}=2\pi\alpha^\prime\left(\frac{1}{g+2\pi\alpha^\prime B}\right)_A^{\mu\nu}~.
\end{flalign}
Note that we extracted the deformation parameter~$\lambda$ on the
left-hand-side in order to obtain a dimensionless~$\Theta^{\mu\nu}$.
Thus, the NC can in principle be related to the dynamics
of the $B$-field using~(\ref{eqn:PoissonBfield}).

To motivate preferred choices for the $y$-dependence of the Poisson tensor we 
work within the supergravity approximation of string theory,
dimensionally reduced to five-dimensional
Anti-de-Sitter~($\text{AdS}_5$) space. 
We briefly comment on the modifications necessary for the case of a RS spacetime.
The dimensional reductions are
obtained via flux-compactifications of the ten-dimensional
supergravities, yielding lower dimensional gauged supergravities.  Of
particular interest are the~$N=4$ and~$N=8$ gauged supergravities on
$\text{AdS}_5$~\cite{Gunaydin:1984qu,Gunaydin:1985cu,Romans:1985ps}.
Both include $B$-fields in the spectrum which satisfy the same type of
`self-dual' equations of motion. Since we are only interested in the
$y$-dependence of the $B$-field, we can for simplicity restrict
ourselves to the~$N=4$ theory discussed in~\cite{Romans:1985ps}. The
$B$-field equation of motion is given by
\begin{flalign}
\label{eqn:Beom}
 \sqrt{\vert g\vert}^{-1} \epsilon^{MNRST} \epsilon^{\alpha\beta}D_R B^\beta_{ST}
    = g_1 \xi^2 B^{\alpha MN}~,
\end{flalign}
where $\alpha,\beta\in\{1,2\}$ label the two $B$-fields of the $N=4$
theory, $M$, $N$, $R$, $S$, and~$T$ are 5 dimensional indices, 
$D_R$ is the covariant derivative and~$\xi$ is the exponentiated dilaton.
The theory admits an AdS$_5$ solution in which $g_1= 2 k$ is related to the 
curvature of the $\text{AdS}_5$ space and~$\xi=1$. 
In order to solve~(\ref{eqn:Beom}) on that background we make the ansatz
\begin{flalign}
 B^{\alpha}_{\mu\nu}(x^\rho,y) = b^{\alpha}_{\mu\nu}~ \zeta^{\alpha}(y)~,\quad
   B^{\alpha}_{5\mu}(x^\rho,y) = 0~,
\end{flalign}
where~$b^{\alpha}_{\mu\nu}$ is a constant antisymmetric matrix, which
we parameterize by `electric' and `magnetic' components, i.e.~we
define $b^\alpha_{0i}=E^\alpha_i$ and
$b^\alpha_{ij}={\epsilon_{ij}}^kB^\alpha_k$.  Plugging this
into~(\ref{eqn:Beom}) we obtain 
\begin{flalign}\label{eq:B-field-zeta}
 \zeta^1(y) = a_1 ~e^{kRy} + a_2~ e^{-kRy}~,\quad \zeta^2(y) = a_1 ~e^{kRy} - a_2 ~e^{-kRy}~, 
\end{flalign}
and $E^2_i=B^1_i$, $B^2_i=-E^1_i$, i.e.~$b^2_{\mu\nu}$ is fixed in
case~$b^1_{\mu\nu}$ is given, or vice versa.  The relevant object
entering the Poisson tensor is a linear combination of the two
$B$-fields\footnote{See~\cite{Lu:1999bw}, where the bosonic sector of
  the~$N=4$ theory was obtained by means of a reduction ansatz from
  the ten-dimensional type IIB supergravity.  The ansatz used there
  shows that the combination~$\hat B+i\hat C_2$, of the
  ten-dimensional NS-NS sector B-field~$\hat B$ and R-R sector gauge
  field~$\hat C_2$, is determined by~$B^1+i B^2$.},
which we denote by
\begin{flalign}\label{eqn:bfield}
  2\pi \alpha^\prime B_{\mu\nu}(y)=a_{+} e^{kRy} b^+_{\mu\nu }
    + a_{-} e^{-kRy} b^{-}_{\mu\nu}~.
\end{flalign}
Note that we have defined for later convenience the parameters~$a_\pm$
and~$b^\pm_{\mu\nu}$ to be dimensionless by extracting the
factor~$2\pi\alpha^\prime$ and that the~$b^{\pm}_{\mu\nu}$ are not
independent.
For the case of a RS spacetime (\ref{eqn:bfield}) is the solution 
on the fundamental domain. The lifting to the covering space is obtained by
periodical continuation with appropriate parity and
fine-tuned brane-localized $B$-field mass terms.

To understand the $y$-dependence of the Poisson
tensor~(\ref{eqn:PoissonBfield}), we consider the special cases
$B^\pm_{\mu\nu}(y):=B_{\mu\nu}(y)\vert_{a_\mp=0}$ separately.
Assuming $\vec{E}_\pm\cdot\vec{B}_\pm=0$, which is a necessary
requirement for an effective field theory
description~\cite{Aharony:2000gz}, we derive
from~(\ref{eqn:PoissonBfield}) the Poisson tensor 
\begin{flalign}
\label{eqn:ThetaPlusminus}
  \lambda \Theta^{\mu\nu}_\pm
    = -2\pi\alpha^\prime
      \frac{a_{\pm} e^{(4\pm 1)k R y}}{a_{\pm}^2 e^{2(2\pm1)k R y}
      (\vec{B}_\pm^2-\vec{E}_\pm^2)+1}\, b_\pm^{\mu\nu}
   \approx
     -\frac{2\pi\alpha^\prime}{a_\pm(\vec{B}_\pm^2-\vec{E}_\pm^2)}
      b^{\mu\nu}_{\pm} ~e^{\mp k R y}~,
\end{flalign}
where $b_\pm^{\mu\nu}:=
\eta^{\mu\rho}\eta^{\nu\sigma}b^\pm_{\rho\sigma}$ is defined via the
flat metric~$\eta^{\mu\nu}$ and the last approximation holds for not
unnaturally small~$a_\pm$ and $\vec{B}_\pm^2-\vec{E}_\pm^2$.  Even
though it was shown in~\cite{Aharony:2000gz} that a unitary NC QFT is
obtained for~$\vec{B}^2\geq \vec{E}^2$, we restrict our discussion to
the case~$\vec{B}^2>\vec{E}^2$, known as space-like NC.

In the following we assume that the Poisson tensor calculated above
will enter as a $\star$-product~(\ref{eqn:RJSproduct}) in the
effective field theory description of the physics in a $B$-field background.  
Note that this is only a physical assumption which we do not derive here
from a more fundamental theory.  The emergence of $\star$-products in
effective field theories due to a constant $B$-field was shown by
Seiberg and Witten for the case of Yang-Mills
theories~\cite{Seiberg:1999vs}.  For the low energy gravity sector of
string theory in a $B$-field background
see~\cite{AlvarezGaume:2006bn}.

Next, we have to set the scales of our effective field theory
description.  Since we have a vanishing dilaton background, we
identify~$\alpha^\prime$ with the five-dimensional Planck length,
$\alpha^\prime=\hat l_P^2$, which will
be~$\mathcal{O}(\text{TeV}^{-2})$.  Due to the exponential
$y$-dependence of 
$\lambda\Theta^{\mu\nu}_\pm$~(\ref{eqn:ThetaPlusminus}), the resulting
$y$-dependent NC scale can be taken to be TeV-scale on one brane,
while being exponentially small on the other brane\footnote{Since the
  Poisson tensor is of dimension (length)$^{-2}$, the effective NC
  scale on the other brane is suppressed roughly by a
  factor~$e^{\frac{1}{2}k R \pi}$.}. 
Because of the exponentially strong localization of the Poisson tensor
towards one of the branes, we can, as a simplifying approximation,
restrict~$\Theta^{\mu\nu}$ to the brane where it is
localized and assume the remaining part of the theory to be
commutative.  We call this the NC-brane approximation.

\section{NC effects on bulk particles}
\label{sec:bulk}
\begin{figure}
\centering
\includegraphics[width=0.42\linewidth]{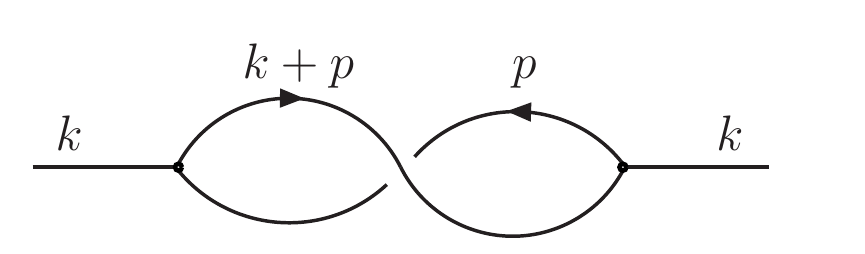}
\caption{\label{fig:selfenergy}Sample diagram for the bulk particle
  self-energies due to NC brane matter. Diagrams of this type induce
  Lorentz-violating effective two-point operators for the bulk
  particle.} 
\end{figure}

Before considering explicit models utilizing the setup described
above, we study the NC effects on bulk particles due to the coupling
to matter on the NC brane in a general context.  We use the
NC-brane approximation throughout this section.
Ultimately, the SM will be confined to the almost commutative brane
and NC effects are communicated from the NC brane to the SM brane via
the bulk particles (e.g.~the KK graviton).  We are thus interested in
the effect of the NC sector on bulk particles, which is the radiative
induction of Lorentz-violating effective operators.  We study these
operators for the radion and the graviton as well as for a bulk U(1)
gauge field at the one-loop level.  For this we first classify all
possible lowest-dimensional Lorentz-violating effective operators
arising from NC interactions and then show that they are indeed
induced via nonplanar self-energies of the type shown in
Fig.~\ref{fig:selfenergy} and simple matter models on the NC brane. In
planar diagrams, the phases arising due to the $\star$-product cancel,
while the nonplanar amplitudes are proportional to~$\cos
(\lambda\,k\Theta p)$.  We expand the integrand in~$\lambda$ up to the
leading order providing NC effects, which is~$\mathcal O(\lambda^2)$.
Kaluza-Klein reduction is not obstructed by the $\star$-product due
to $\starcom{x^\mu}{y}=0$, and is applied throughout.

\paragraph{Radion and graviton}
Applying a graviton expansion to the five-dimensional Einstein Hilbert
action on the $\text{AdS}_5$ background with two branes and using
suitable gauge fixings to extract and decouple the physical degrees of
freedom, the Kaluza-Klein expanded gravity action
reads~\cite{Boos:2002ik,Boos:2002hf}
\begin{subequations}
\label{eqn:bulkaction}
\begin{flalign}
S_\text{bulk}&=\frac{1}{2}\int d^4z\, \partial_\mu\phi\partial^\mu\phi 
+ \frac{1}{4}\sum_n\int d^4z\left( \partial_\mu
  u_{\nu\rho}^{(n)}\partial^\mu {u^{(n)}}^{\nu\rho}-m_n^2
  e^{2kR\pi}u^{(n)}_{\nu\rho}{u^{(n)}}^{\nu\rho}\right)~,\\ 
S_\text{NC-brane}^\text{int}&=
\sum_n\frac{\kappa_n}{2}\int
  d^4z\  u_{\mu\nu}^{(n)}T_\star^{\mu\nu}+\frac{\kappa_\phi}{2\sqrt{3}}\int
  d^4z\, \phi {T_\star^\mu}_\mu~,
\end{flalign}
\end{subequations}
where~$\kappa_n$ and~$\kappa_\phi$ are coupling constants of
dimension~$-1$, which are related to the five-dimensional Planck mass,
overlaps of the KK gravitons and the radion with the NC brane and
exponential factors~$e^{k R\pi}$. These relations will be discussed in
detail later. $z^\mu$ denotes coordinates which are Galilean on the SM
brane, i.e.~$z^\mu\!=\!e^{-kR\pi}x^\mu$,~$\phi$ denotes the radion and~$u^{(n)}_{\mu\nu}$ are the
graviton KK modes.  By cyclicity of the integral, in each
term one of the~$\star$-products can be replaced by an ordinary
product.  Thus the $\star$-products drop out of the quadratic part of
the action and all NC effects can be absorbed into the definition of
the deformed stress-energy tensor~$T_\star^{\mu\nu}$ on the NC brane.

We now consider the lowest-dimensional Lorentz-violating effective
operators for the radion field. Assuming for simplicity massless
matter fields on the NC brane, we obtain 
\begin{flalign}\label{eq:radion-ct}
 S^\text{s}_\text{eff}=\frac{(\kappa_\phi\lambda)^2}{2}
   \int d^4z\, a_1^{\text{s}}\phi \tilde\partial^2\partial^6\phi~,
\end{flalign}
where~$(\kappa_\phi\lambda)^2$ has dimension~$-6$, $a_1^{\text{s}}$ is
dimensionless and $\tilde\partial^\mu:=\Theta^{\mu\nu}\partial_\nu$. 
Note that, if we had considered massive fields on the NC brane, the
operator as given above with any of the~$\partial^2$ replaced by the
squared mass would also be possible. 
To show by an explicit example that this operator is indeed induced,
we consider an abelian gauge field~$A_\mu$ and a massless real
scalar~$\chi$ on the NC brane, described by 
\begin{subequations}
\label{eqn:NCbraneAction}
\begin{flalign}
S_\text{NC-brane}&=-\frac{1}{4}\int d^4z
  F^{\mu\nu}F_{\mu\nu}+\frac{1}{2}\int
  d^4z\partial_\mu\chi\partial^\mu\chi~,\\ 
T_\star^{\mu\nu}&=F^{\mu\rho}\star
  {F^{\nu}}_\rho-\frac{1}{4}\eta^{\mu\nu}F^{\rho\sigma}\star
  F_{\rho\sigma}-\partial^\mu\chi\star\partial^\nu\chi
  +\frac{\eta^{\mu\nu}}{2}\partial_\rho\chi\star\partial^\rho\chi~. 
\end{flalign}
\end{subequations}
We calculate the nonplanar radion self-energy~$\Pi(k)$ due to the
NC-brane scalar field (see Fig.~\ref{fig:selfenergy}) using dimensional
regularization and find $\Pi(k)= a_1^{\text{s}} \frac{i}{\epsilon}\kappa_\phi^2
\lambda^2 k^6\tilde k^2+\text{finite}$, with~$a_1^{\text{s}}\not=0$.  
This shows that the expected operator is indeed induced for a simple
matter model on the NC brane. The self-energy due to the NC-brane
vector field vanishes, since the corresponding part of the
stress-energy tensor is traceless. 

Next, we consider the graviton.  The four-dimensional gauge
transformations of the massless graviton mode~$u_{\mu\nu}^{(0)}$ are
given by
\begin{flalign}
u_{\mu\nu}^{(0)}\rightarrow
  {u_{\mu\nu}^{(0)}}^\prime=u_{\mu\nu}^{(0)}-\partial_\mu
  \zeta_\nu-\partial_\nu\zeta_\mu~. 
\end{flalign}
Noting that both $\tilde\partial^\mu\tilde\partial^\nu
u_{\mu\nu}^{(0)}$ and the linearized scalar curvature
$R=\kappa_0\left(\eta^{\mu\nu}\partial^\rho\partial_\rho
-\partial^\mu\partial^\nu\right) u^{(0)}_{\mu\nu}+\mathcal
O(\kappa_0^2)=:\kappa_0 D^{\mu\nu}u^{(0)}_{\mu\nu}+\mathcal
O(\kappa_0^2)$ are gauge invariant, we can write the following gauge
invariant and Lorentz-violating effective operators for the graviton
\begin{multline}\label{eq:graviton-operators}
\mathcal S_\text{eff}^\text{grav}=
  \frac{\left(\kappa_n\lambda\right)^{2}}{4}\int d^4z\, 
  u^{(n)}_{\mu\nu}  \Bigl(
      c_1\, D^{\mu\nu} \tilde\partial^\rho\tilde\partial^\sigma \partial^2 \\
    + c_2\, D^{\mu\rho}\tilde\partial^\nu \tilde\partial^\sigma  \partial^2 
    + c_3\, D^{\mu\nu} D^{\rho\sigma} \tilde\partial^2
    + c_4\, D^{\mu\rho}D^{\nu\sigma}\tilde\partial^2
  \Bigr)\partial^2u^{(n)}_{\rho\sigma}~.  
\end{multline}
These are all possible effective operators, because in one-loop
diagrams the Poisson tensor is always contracted with one external
momentum.  Furthermore, four-dimensional gauge invariance restricts
the possible loop-induced operators for all graviton modes, not just
for the massless mode.  Indeed, the structure of the couplings is the
same for massless and massive modes. Therefore, if an effective
operator for a higher KK graviton is induced by a self-energy diagram
as in Fig.~\ref{fig:selfenergy}, the same divergence appears in the
massless graviton self-energy. Consequently, the effective
graviton operators are restricted to those
of~(\ref{eq:graviton-operators}) for all KK modes.

Explicitly calculating the nonplanar graviton self-energies
$\Pi_\text{s}^{\mu\nu\rho\sigma}(k)$
and~$\Pi_\text{v}^{\mu\nu\rho\sigma}(k)$ in the simple matter model of
a NC-brane scalar and vector field, respectively, we find\footnote{%
  The brane-localized interactions introduce mixing between the KK
  modes which, however, will be ignored in the following since it is 
  not relevant to our investigations.}
\begin{equation}\label{eq:graviton-selfenergy}
  \Pi_\text{s/v}^{\mu\nu\rho\sigma}(k) =
  \frac{i}{\epsilon}\kappa_n^2\lambda^2\sum_{i=1}^4c_i^\text{s/v}
     O_i^{\mu\nu\rho\sigma}(k)+\text{finite}~, 
\end{equation}
where all the~$c_i^\text{s}$ and~$c_i^\text{v}$ are nonzero
and~$O_i^{\mu\nu\rho\sigma}(k)$ are the momentum-space versions of the
operators in~(\ref{eq:graviton-operators}), as given in
App.~\ref{app:feynmanrules}.  Hence, all gauge invariant
Lorentz-violating effective operators given above are induced
radiatively.  We find that the $\Pi_\text{s,v}^{\mu\nu\rho\sigma}(k)$
fulfill the standard Ward identity
$k_\mu\Pi_\text{s,v}^{\mu\nu\rho\sigma}(k)=0$ (and the ones arising
from permutations of the indices of
$\Pi_\text{s,v}^{\mu\nu\rho\sigma}(k)$), which can be understood from
the fact that the action~(\ref{eqn:bulkaction}) consists of free
gravitons coupled to a conserved $\star$-deformed matter
current~$T_\star^{\mu\nu}$.  Whether these Ward identities
continue to hold at higher orders in the gravitational coupling, or
whether they have to be deformed as well, is still an open question,
because it is in general not possible to absorb all NC effects into a
deformed matter stress-energy tensor~$T_\star^{\mu\nu}$ in this case.

\paragraph{Abelian gauge field}
We now turn to a bulk U(1) gauge field~$A_{M}$ ($M=0,1,2,3,5$), which
we assume to be coupled to a charged scalar~$\chi$ on the NC brane
(the fields~$A_\mu$ and~$\chi$ are unrelated to those of the previous
paragraph).  After gauge fixing and Kaluza-Klein expansion
$A_{\mu/5}(z,y)=\sum_n A_{\mu/5}^{(n)}(z)\, t_{1/2}^{(n)}(y)$ the action is given by
\begin{subequations}\label{bulk-vector-action}
\begin{eqnarray}
S_\text{bulk}
&=&
\frac{1}{2}\sum_n\int d^4z
A_\mu^{(n)}
\left(
\left(\partial^2+m_n^2\right)\eta^{\mu\nu}-\left(1-\frac{1}{\xi}\right)\partial^\mu\partial^\nu
\right)
A_\nu^{(n)}\nonumber\\
&&
-\frac{1}{2}\sum_n\int d^4z A_5^{(n)}\left(\partial^2+\xi m_n^2\right)A_5^{(n)}~,\\
S_\text{NC-brane}&=&\int d^4z (D_\mu\chi)^\dagger\star D^\mu\chi~,\qquad
  D_\mu\chi=\partial_\mu\chi+i\sum_n g_n A_\mu^{(n)}\star\chi~.\label{eqn:BulkGaugeField-int}
\end{eqnarray}
\end{subequations}
We choose Neumann/Dirichlet boundary conditions for~$A_\mu$/$A_5$, and
apply unitary gauge to decouple the massive scalar modes.  The
dimensionless coupling constants~$g_n$ contain the gauge
coupling~$g_5$ and the brane-overlap $t_1^{(n)}\!(y_\text{{\tiny NC-brane}})$.
The four-dimensional gauge transformations are given by
$A^{(0)}_\mu\rightarrow A^{(0)}_\mu+\partial_\mu\omega$ and, as
explained above, restrict the possible loop-induced effective operators
for all KK modes.  Using the restrictions from gauge invariance, we
find the following two independent Lorentz-violating effective operators
\begin{flalign}\label{eqn:vector-ct}
 S^\text{A}_\text{eff}=
\frac{(g_n\lambda)^{2}}{4}\int d^4z\,F^{(n)}_{\mu\nu}\Bigl(
  d_1\Theta^{\mu\rho}\Theta^{\nu\sigma}\partial^4
  + d_2 \eta^{\mu\rho}\eta^{\nu\sigma}\tilde\partial^2\partial^2\Bigr)
   F^{(n)}_{\rho\sigma}~,
\end{flalign}
where $F^{(n)}_{\mu\nu}=\partial_{[\mu}A^{(n)}_{\nu]}$. 

Using the simple NC brane matter model~(\ref{eqn:BulkGaugeField-int}),
we find for the nonplanar gauge boson self-energy
$\Pi^{\mu\nu}(k)=\frac{i}{\epsilon}g_n^2\lambda^2\left(d_1
O_1^{\mu\nu}(k)+d_2O_2^{\mu\nu}(k)\right)+\text{finite}$,
where~$O_i^{\mu\nu}(k)$ are the momentum-space versions of the
operators given in~(\ref{eqn:vector-ct}) and~$d_1,d_2\neq0$.  This
shows that both Lorentz-violating effective operators are induced at
the one-loop level.

\section{Explicit models}
\label{sec:models}
We now discuss two explicit models which realize the idea of
transmitting NC effects via bulk particles from a NC sector to the SM.
In both models the SM is confined to the IR brane where the $B$-field
solution is chosen such that spacetime NC is
exponentially suppressed.  As explained in Section \ref{sec:stringNC},
we apply the NC-brane approximation after constructing the models and
assume the bulk and the SM brane to be commutative.  The general
effects of a NC brane on bulk particles have been discussed in the
previous section, so the remaining task is to determine the coupling
strengths of the bulk particles to the different branes.  While the
localizations of the KK gravitons and the radion are fixed by the
geometry, leading to fixed coupling constants~$\kappa_\phi$
and~$\kappa_{n}$ in~(\ref{eqn:bulkaction}), the bulk gauge field
offers additional freedom to fix its localization, e.g.~via boundary
kinetic/mass terms.

\begin{figure}
\center
\subfigure[][]{
\label{fig:modelpicture-a}
\includegraphics[width=0.21\linewidth]{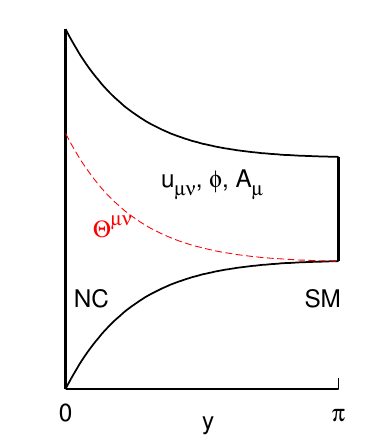}
}\qquad
\subfigure[][]{
\label{fig:modelpicture-b}
\includegraphics[width=0.35\linewidth]{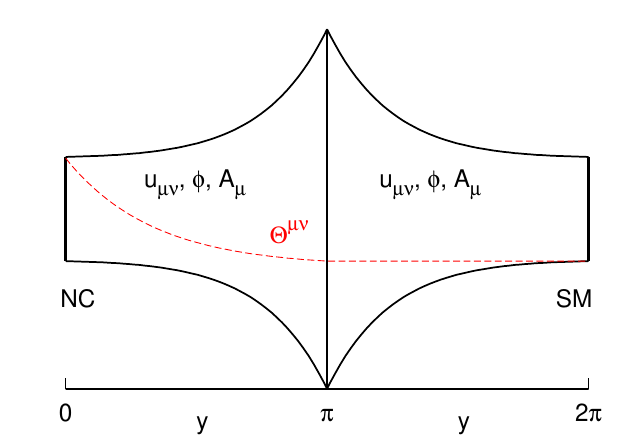}}
\caption{\label{fig:modelpicture}The basic setups for our two
  models. \subref{fig:modelpicture-a} shows the standard RS setup with
  NC UV brane, \subref{fig:modelpicture-b} shows the
  three-brane model with the second NC IR brane.}
\end{figure}

The setup for the two models we are considering is depicted in
Fig.~\ref{fig:modelpicture}.  In the first part we discuss the
standard RS setup with a UV-brane localized NC (see
Fig.~\ref{fig:modelpicture-a}) and in the second part we use an
extended RS scenario, where the NC is localized on a second IR
brane (see Fig.~\ref{fig:modelpicture-b}).

\subsection{NC Randall-Sundrum model}
We consider the standard RS setup shown in
Fig.~\ref{fig:modelpicture-a}.  The $B$-field configuration is chosen
as~$B^+_{\mu\nu}$, implying the Poisson tensor~$\Theta^{\mu\nu}_+$
of~(\ref{eqn:ThetaPlusminus}), such that the IR brane, where the SM is
localized, is almost commutative, while the UV brane is not.  To
determine the brane couplings of the KK gravitons and radion we
require their mode functions entering the KK reduction of the
five-dimensional action.  The KK ansatz and the mode functions are
given by~\cite{Boos:2002ik,Boos:2002hf}
\begin{subequations}
\begin{flalign}
&u_{\mu\nu}(x,y)=\sum_n u_{\mu\nu}^{(n)}(x) f_n(y)~,\qquad\qquad \int
  dy\,e^{2kR|y|}f_m(y)f_n(y)=\delta_{nm}~,\\ 
&f_0(y)=N_0 e^{-2k R|y|}~,\qquad
  f_n(y)=N_n\left(Y_1\left(\gamma_n\right)J_2(\gamma_n
  e^{kR|y|})-J_1\left(\gamma_n\right)Y_2(\gamma_n
  e^{kR|y|})\right)~,\label{eqn:graviton-modes} 
\end{flalign}
\end{subequations}
where  $J_1(\gamma_n e^{kR\pi})=0$ and~$m_n=\gamma_n k$.

The coupling of gravity to the SM brane at~$y=\pi$ is given
by~(\ref{eqn:bulkaction}) with
$\kappa_{n/\phi}=\kappa_{n/\phi}^\text{SM}$.  We define
$\kappa_\text{SM}:=\kappa_0^\text{SM}$ as the coupling of the massless
graviton mode to SM matter, which has to be fixed to
$\kappa_\text{SM}=\sqrt{16\pi} M_\text{Planck}^{-1}$ to reproduce the
observed four-dimensional gravitational coupling. One obtains the
relations
\begin{flalign}\label{eqn:SMcouplingsRS1}
\kappa_0^\text{SM}=\hat\kappa f_0(\pi)e^{kR\pi}~,
\qquad\kappa_n^\text{SM}=\kappa_\text{SM}\frac{f_n(\pi)}{f_0(\pi)}~, 
\qquad\kappa_\phi^\text{SM}=\kappa_\text{SM}e^{kR\pi}~.
\end{flalign}
The five-dimensional gravitational coupling~$\hat\kappa$ is chosen
such that $\kappa_\text{SM}=\sqrt{16\pi}M_\text{Planck}^{-1}$,
i.e.~$\hat\kappa=\kappa_\text{SM} e^{kR\pi}/N_0$.  For a typical value
of~$kR\approx11$ the first KK mode of the graviton has a mass of order
TeV and a SM-brane coupling of order $\textrm{TeV}^{-1}$.  The radion
couples to the SM brane with a similar strength as the massive
gravitons.

The coupling to the UV brane, localized at~$y=0$, is given
by~(\ref{eqn:bulkaction}) with
$\kappa_{n/\phi}=\kappa_{n/\phi}^\text{UV}$, where
\begin{flalign}\label{eqn:UVcouplingsRS1}
\kappa_n^\text{UV}=\hat\kappa
f_n(0)e^{-kR\pi},\qquad\kappa^\text{UV}_\phi=\kappa_\text{SM}e^{-kR\pi}~. 
\end{flalign}
Note that this is different from the couplings derived
in~\cite{Boos:2002ik,Boos:2002hf} for a UV-brane observer, since we
describe the UV-brane physics as seen from the IR brane, inducing the
additional redshift factor~$e^{-k R \pi}$.

The massive gravitons are of phenomenological interest due to their
enhanced couplings to the SM brane.  However, from
expressions~(\ref{eqn:SMcouplingsRS1}) and~(\ref{eqn:UVcouplingsRS1})
we find that the coupling of the massive gravitons to the NC brane is
exponentially suppressed.  Thus, the Lorentz-violating
operators~(\ref{eq:graviton-operators}) are Planck-scale suppressed
and their effect for SM physics at the TeV scale is negligible. The
same applies to the radion.

In contrast to the graviton and radion where the boundary conditions
are fixed by the presence of the branes, the localization of a bulk
vector field can be adjusted by a suitable choice of boundary
conditions and brane-localized mass and/or kinetic terms.  We will
therefore employ an abelian bulk gauge field here and realize
observable NC effects via the graviton in the next section in an
extended RS setup.  To the bulk and NC-brane actions
of~(\ref{bulk-vector-action}) we add the IR-brane action
\begin{equation}
S_\text{IR-brane}=\int d^4 z\left( D^\mu\phi
D_\mu\phi+V(\phi)+\mathcal L_\text{SM}\right),\qquad
D_\mu\phi=\partial_\mu\phi+i \sum_n g_5 A_\mu^{(n)} t_1^{(n)}\!(\pi)\phi ~,
\end{equation}
where~$\mathcal L_\text{SM}$ denotes the Lagrangian of the SM and possible
interactions with the bulk gauge field.  The scalar~$\phi$ is charged
under the bulk U(1) and~$V(\phi)$ is such that $\left<\phi\right>=v$
spontaneously breaks the bulk gauge symmetry. We then find the brane-localized mass term
$-g_5^2 v^2 A_\mu^{(n)}
A_\nu^{(m)}\eta^{\mu\nu} t_1^{(n)}\!(\pi) t_1^{(m)}\!(\pi)\delta(y-\pi)$.  With Neumann boundary
condition on the NC brane the KK wave functions~$t_1^{(n)}$ of~$A_\mu^{(n)}$
are given by
\begin{equation}
t_1^{(n)}\!(y)=
N_n e^{k R (y-\pi)}\left( J_1\left(\frac{m}{k} e^{k R (y-\pi)}\right)
Y_0\left(\frac{m}{k}\right)
- 
Y_1\left(\frac{m}{k} e^{k R (y-\pi)}\right)
J_0\left(\frac{m}{k}\right)\right)~.
\end{equation}
Focussing on the lightest KK mode, we find 
from the IR-brane boundary condition ${t_1^{(n)}}^\prime-2 R g_5^2 v^2
t_1^{(n)}\vert_{y=\pi}=0$ a mass of order
TeV. Furthermore, the lightest mode is almost delocalized, with appreciable coupling to both branes.
It can therefore transmit NC effects to the SM brane.

\subsection{Extended NC Randall-Sundrum model}
The second model realizes NC effects via the KK gravitons. The basic
setup is depicted in Fig.~\ref{fig:modelpicture-b}.  As in the
standard RS setup we consider the orbifold $\mathbb R^{1,3}\times
S^1/\mathbb Z_2$.  We parameterize the circle by~$y$ running
from~$-2\pi$ to~$2\pi$.
Two branes of tension~$\Lambda_1$, $\Lambda_3$ are placed at the
orbifold fixed-points~$y_1=0$, $y_3=2\pi$ and a third one of
tension~$\Lambda_2$ at~$y_2=\pi$.  We solve the associated Einstein
equations
\begin{flalign}
R_{MN}-\frac{1}{2}g_{MN}R
= -\frac{1}{4M_5^3}\left(\Lambda g_{MN}+\sum_{i=1}^3
\Lambda_i\delta(y-y_i) \frac{\sqrt{-\hat g^{(i)}}}{\sqrt{g}}\hat
g^{(i)}_{\mu\nu}\delta^\mu_M\delta^\nu_N\right)~,
\end{flalign}
where~$\hat g^{(i)}$ denotes the induced metric on the $i$-th brane.
With the ansatz $ds^2=e^{-2\sigma(\hat y)}\eta_{\mu\nu}dx^\mu\otimes
dx^\nu-R^2dy\otimes dy$, where $\hat y(y)=||y|-\pi|$ is consistent
with the orbifold symmetry, we find the solution~$\sigma(\hat y)=k R \hat y$,
where~$k$ is defined by~$\Lambda=-24 M_5^3 k^2$,
and~$\Lambda_1=\Lambda_3=\Lambda/k$, $\Lambda_2=-\Lambda_1$.  We thus
find two standard RS backgrounds glued together at the UV brane.

The $B$-field configuration is chosen as~$B^-_{\mu\nu}(\pi-y)$
for~$y\in[0,\pi]$ and~$B^+_{\mu\nu}(y-\pi)$ for~$y\in(\pi,2\pi]$, such
that~$B$ is continuous at the UV brane at~$y=\pi$.  By this choice,
we obtain the continuous Poisson tensor~$\Theta^{\mu\nu}_-(\pi-y)$
for~$y\in[0,\pi]$ and $\Theta^{\mu\nu}_+(y-\pi)$
for~$y\in(\pi,2\pi]$, see~(\ref{eqn:ThetaPlusminus}), which exponentially
grows towards the IR brane at~$y=0$.  This yields in the NC-brane
approximation a commutative IR brane at~$y=2\pi$ where the SM is
located, a commutative UV brane at~$y=\pi$, and a NC
IR brane at~$y=0$.

Performing the graviton expansion $u_{\mu\nu}(x,y)=\sum_n
u_{\mu\nu}^{(n)}(x) \hat f_n(y)$, we find $\hat f_n(y)=f_n(\hat y)$
with~$f_n$ given in~(\ref{eqn:graviton-modes}).  The masses of the
graviton modes and their matter couplings on the SM and UV brane are
equal to those of the standard RS case.  The matter couplings on the
NC brane are equal to those on the SM brane, i.e.~of order TeV$^{-1}$
for the massive modes.  Thus, the phenomenology of this setup is very
similar to that of the standard RS scenario, but we do have an
effective communication of Lorentz-violating effects from the NC brane
to the SM brane via gravity.

\section{Options for phenomenology}
\label{sec:pheno}
In the previous section we have constructed two models with an
effective transmission of Lorentz violation from the NC brane to the
SM brane.  The main difference is the messenger particle, which is a
massive vector field in the first one and the massive graviton in the
second.  As an illustration, we discuss in this section an exemplary
collider process and a way to compare our setup to experimental
constraints on the NC scale.

\subsection{NC effects in scattering processes}
\begin{figure}
 \centering
 \setbox1=\hbox{\includegraphics[width=0.3\linewidth]{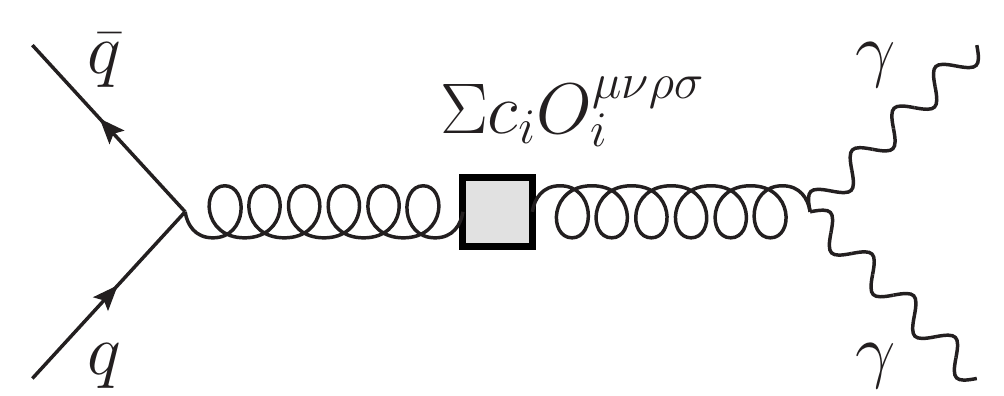}}
 \setbox2=\hbox{\includegraphics[width=0.32\linewidth]{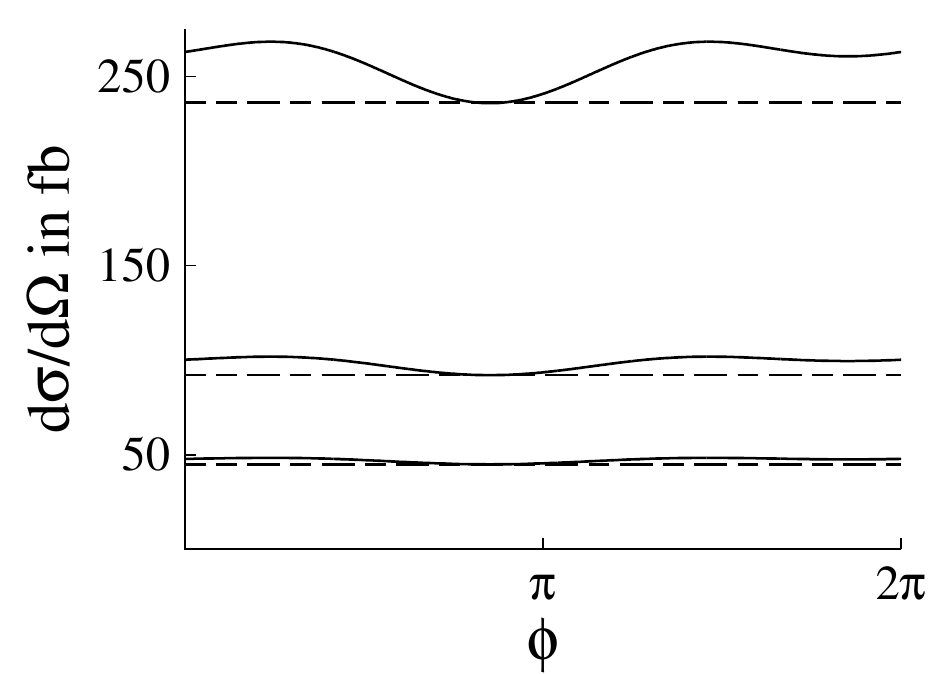}}
 \subfigure[][]{
    \label{fig:process-diagram}
    \raisebox{0.5\ht2-0.5\ht1}{
      \includegraphics[width=0.3\linewidth]{process}
    }
  }\qquad\qquad
  \subfigure[][]{
    \label{fig:plot}
    \includegraphics[width=0.32\linewidth]{qq-asym}}
 \caption{\label{fig:diagramplot} \subref{fig:process-diagram}
   Lorentz-violating contribution to $\bar q q\rightarrow
   \gamma\gamma$ in the three-brane model. \subref{fig:plot} Cross
   section for $\bar q q\rightarrow \gamma\gamma$ scattering with a
   polar angle of~$\pi/3$ and $\sqrt{s}\in\lbrace
   0.985,0.990,0.995\rbrace~\text{TeV}$. Dashed lines show the cross
   section for the commutative case~$\lambda=0$, solid lines for~$\lambda=0.1~\text{TeV}^{-2}$, 
   see~(\ref{parameters-cross-section}). The SM contributions are
   ignored, since they are subdominant at the resonance.}
\end{figure}

We consider the scattering of two massless quarks~$q$, producing two
SM photons~$\gamma$ in the extended RS model.  In addition to the
s-channel exchange of the massive graviton, $\bar{q}q\rightarrow
u_{\mu\nu}^{(n)}\rightarrow\gamma\gamma$, we have
the~$\mathcal{O}(\lambda^2)$ Lorentz-violating contributions shown in
Fig.~\ref{fig:process-diagram}.

The relevant model parameters are the mass and width of the first KK
graviton mode and its coupling to the SM and NC brane.  We fix them by
the reasonable choice~\cite{Davoudiasl:2000wi}
\begin{flalign}\label{parameters-cross-section}
 m_1=1~\text{TeV}~,\qquad\kappa_1^{SM}=0.1
 \beta_1~\text{TeV}^{-1}~,\qquad\Gamma_1=0.01~\text{TeV}~,\qquad
 \lambda =0.1~\text{TeV}^{-2} ~,
\end{flalign}
where~$\beta_1=3.83$ is the first root of~$J_1$.

Due to the Lorentz-violating operators, the cross section shows the
typical azimuthal-angle dependence, as shown in Fig.~\ref{fig:plot}.
This dependence is due to the~$\vec{E}$ components of the Poisson
tensor.  We find a similar effect in the scattering of vector bosons,
and also in the standard RS model for cross sections involving the Lorentz-violating
contribution from the massive bulk vector field.  The main feature of
our models is the amplification of the NC effects at the resonance,
while they are strongly suppressed outside the resonance region.  This
offers the possibility to detect -- together with the additional
massive particles due to the extradimensional setup -- effects of
spacetime NC in particle collisions at the TeV scale.

\subsection{NC effects on the SM photon}
Comparing our models to experimental bounds from laboratory
experiments and astrophysics is possible by considering the effective
Lorentz-violating operators which are induced for the SM photon
propagator.  In our models, such operators are induced by the photon
self-energy due to the massive bulk particle with effective-operator
insertions.  Hence, they effectively are an (at least) two-loop effect
and there is no resonance enhancement, implying strong suppression
compared to the effect on scattering cross sections.

\begin{figure}
\centering
\includegraphics[width=0.3\linewidth]{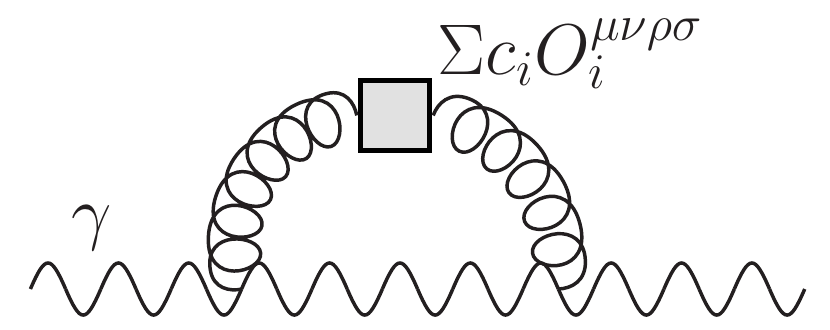}\qquad
\includegraphics[width=0.3\linewidth]{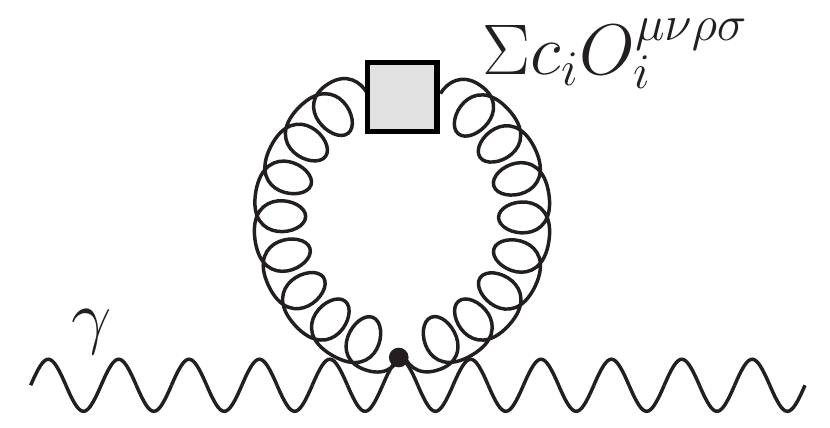}
\caption{\label{fig:photon-selfenergy}NC contributions to the photon
  self-energy in the extended RS model.} 
\end{figure}

We first discuss the second model, where the graviton mediates
NC and the photon is affected by the presence of NC
interactions by the self-energy diagrams shown in
Fig.~\ref{fig:photon-selfenergy}.  The Lorentz-violating operators at
lowest order~$\mathcal O(\lambda^2)$ are
\begin{flalign}\label{eqn:photon-operators}
S^\gamma_\text{eff}=
\kappa_n^4\lambda^{2}
\int d^4z\,F_{\mu\nu}\Bigl(
d_1^\gamma\,\Theta^{\mu\rho}\Theta^{\nu\sigma}m_n^8 
+ d_2^\gamma\, \eta^{\mu\rho}\eta^{\nu\sigma}m_n^6 \tilde\partial^2
+ d_3^\gamma\, \Theta^{\mu\lambda}{\Theta_\lambda}^\rho \eta^{\nu\sigma}m_n^8
\Bigr) F_{\rho\sigma}~,
\end{flalign}
and those with~$m_n^2$ replaced by~$\partial^2$. 
However, the relevant operators for laboratory experiments at low
energy and for astrophysics, where photons are on-shell, are those
with a minimal number of derivatives.  We calculate the
Lorentz-violating divergent part of the diagrams in
Fig.~\ref{fig:photon-selfenergy}, i.e.~the $\beta$-functions of
the~$d_i^\gamma$, and find it proportional to the third operator
in~(\ref{eqn:photon-operators}) and $3c_1+17c_2-45c_4$.  The
coefficient~$c_3$ does not appear since,
with~$B_{\mu\nu\rho\sigma}(p)$ denoting the numerator of the massive
graviton propagator, $B_{\mu\nu\mu^\prime\nu^\prime}(p)
O_3^{\mu^\prime\nu^\prime\rho^\prime\sigma^\prime}(p)
B_{\rho^\prime\sigma^\prime\rho\sigma}(p)\propto (p^2-m^2)^2$ cancels
the two massive propagator denominators, leaving only scale-free
integrals.  This linear combination of the~$c_i$ can be chosen to
vanish, such that~$d_i^\gamma=0$ can be fixed at laboratory energy
scales and is preserved by renormalization group running at leading
order.  We note that this does not affect the azimuthal-angle
dependence in the scattering cross section, which depends on~$c_2$
only.  A detailed discussion of experimental bounds on the entries
of~$\kappa^{\mu\nu\rho\sigma}$ in
$-\frac{1}{4}\kappa^{\mu\nu\rho\sigma} F_{\mu\nu}F_{\rho\sigma}$ can
be found in~\cite{Kostelecky:2002hh,Kostelecky:2008ts}.

In the first model, the situation is even better.  Since the SM photon
is not charged under the bulk U(1) gauge group, it is affected
starting effectively at the three-loop order, and we expect an
additional suppression in that case.

\section{Conclusion}
\label{sec:concl}
We have studied deformations of the Randall-Sundrum spacetime induced
by a Poisson tensor depending on the radial coordinate~$y$, and
constructed a semi-Killing $\star$-product realizing the associated
commutation relations.  Due to the semi-Killing property, the
classical Randall-Sundrum metric solves the NC Einstein equations for
generic $y$-dependence.  Exploiting the relation of the Poisson tensor
to the B-field as motivated by string theory in the low-energy
description in terms of gauged supergravity on $\text{AdS}_5$, we
found preferred choices for the radial-coordinate dependence where the
Poisson tensor is exponentially localized towards either the UV or IR
brane.

Assuming the Poisson tensor to be confined to one of the branes, we
classified the effective Lorentz-violating operators arising from
brane-localized NC interactions for bulk particles, namely the
graviton, radion and a bulk U(1) gauge field.  Choosing simple matter
models on the NC brane, we have shown that these operators are induced
radiatively.  We then constructed two models with an effective
transmission of NC effects to an (almost) commutative SM brane via
bulk particles.  In the standard RS setup we had to employ a bulk
gauge field, while in an extended model with three branes the massive
gravitons are well suited.

As a phenomenological illustration we gave a scattering cross section
which -- close to the bulk-particle resonance -- shows the typical
azimuthal-angle dependence, and briefly studied the effective
operators induced for the SM photon.  This calls for more detailed
model building and
phenomenological studies, which, however, is beyond the scope of this
work.

\section*{Acknowledgements}
We thank Paolo Aschieri and Thomas Flacke for valuable discussions and
comments.  
AS is supported by Deutsche
Forschungsgemeinschaft through the Research Training Group GRK\,1147 
\textit{Theoretical Astrophysics and Particle Physics}.
CFU is supported by the German National Academic Foundation 
(Studienstiftung des deutschen Volkes) and by Deutsche
Forschungsgemeinschaft through the Research Training Group GRK\,1147 
\textit{Theoretical Astrophysics and Particle Physics}.

\appendix
\section{Feynman rules\label{app:feynmanrules}}
For explicitness we list the additional $\mathcal O(\lambda^2)$ Feynman rules 
for the bulk-particle KK modes and give the
momentum-space expressions of the effective two-point
operators~$O_i^{\mu\nu\rho\sigma}(k)$ for the graviton
and~$O_i^{\mu\nu}(k)$ for the bulk gauge field:
\begin{subequations}
\begin{flalign}
u_{\mu\nu}^{(n)}:\quad&
\quad\text{\includegraphics[width=3.3cm]{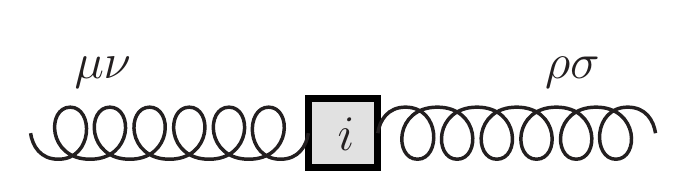}}\qquad
i\kappa_n^2\,\lambda^2\, c_i\, O^{\mu\nu\rho\sigma}_i(k)~,\\ 
\phi:\quad& \quad\text{\includegraphics[width=3.3cm]{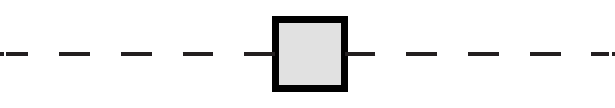}}\qquad
i\kappa_\phi^2\,\lambda^2 \,a_1 \tilde k^2\,k^6~,\\ 
A_\mu^{(n)}:\quad& \quad
\text{\includegraphics[width=3.3cm]{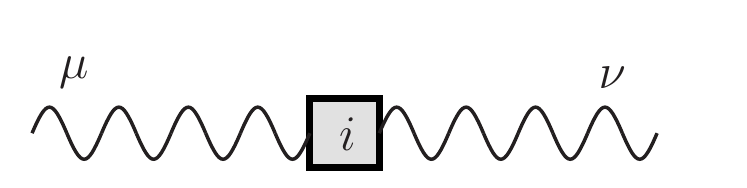}}\qquad i g_n^2\,
\lambda^2\, d_i \,O_i^{\mu\nu}(k)~. 
\end{flalign}
\end{subequations}
The effective operators are $O_1^{\mu\nu}(k)=k^4\,\tilde k^\mu \,\tilde k^\nu$,\; 
$O_2^{\mu\nu}(k)=\tilde k^2\,k^2\,\left(k^\mu k^\nu-\eta^{\mu\nu}k^2\right)$ and
\begin{subequations}
\label{eqn:LVoperator}
\begin{flalign}
O_1^{\mu\nu\rho\sigma}(k) &=\text{Sym}\left[k^4 \,(k^\mu k^\nu -
  \eta^{\mu\nu} k^2)\tilde k^\rho \tilde k^\sigma \right] ~, \\ 
O_2^{\mu\nu\rho\sigma}(k) &=\text{Sym}\left[k^4 \,(k^\mu k^\rho -
  \eta^{\mu\rho} k^2)\tilde k^\nu \tilde k^\sigma  \right] ~,\\ 
O_3^{\mu\nu\rho\sigma}(k) &=\tilde k^2\,k^2\,(k^\mu k^\nu
-\eta^{\mu\nu}k^2)~(k^\rho k^\sigma - \eta^{\rho\sigma} k^2)~,\\ 
O_4^{\mu\nu\rho\sigma}(k) &=\text{Sym}\left[\tilde k^2\,k^2\,(k^\mu
  k^\rho -\eta^{\mu\rho}k^2)~(k^\nu k^\sigma - \eta^{\nu\sigma}
  k^2)\right]~, 
\end{flalign}
\end{subequations}
where `Sym' indicates normalized symmetrization in the index
pairs~$\mu\nu$ and~$\rho\sigma$, and in
$\mu\nu\leftrightarrow\rho\sigma$.

\providecommand{\href}[2]{#2}
\begingroup\raggedright
\endgroup

\end{document}